\documentclass[9pt,shortpaper,twoside,web]{ieeecolor2}

\usepackage{generic}
\usepackage{amsmath,amssymb,amsfonts}
\usepackage{algorithm}
\usepackage{algorithmic}
\usepackage{graphicx}
\usepackage{fancyhdr}
\usepackage{textcomp}
\usepackage{booktabs}
\usepackage{array}
\usepackage{tabularx}
\usepackage{makecell}

\makeatletter
\let\NAT@parse\undefined
\makeatother

\usepackage{hyperref}
\hypersetup{
    colorlinks=true,
    citecolor=blue,
    linkcolor=blue,
    urlcolor=blue
}

\def\BibTeX{{\rm B\kern-.05em{\sc i\kern-.025em b}\kern-.08em
    T\kern-.1667em\lower.7ex\hbox{E}\kern-.125emX}}

\begin{document}

\title{MoWaveQFormer: A Motion-Conditioned Quality-Gated Transformer for Smartphone-Based PPG Heart Rate Estimation}

\author{Swandip Singha, Rifat Bin Reza, and Saifur Rahman Sabuj
\thanks{Swandip Singha is with the Department of Electrical and Electronic Engineering, Chittagong University of Engineering and Technology, Chattogram, Bangladesh (e-mail: u2102181@student.cuet.ac.bd).}%
\thanks{Rifat Bin Reza is with the Department of Electrical and Electronic Engineering, BRAC University, Dhaka, Bangladesh \\(e-mail: rifat.bin.reza1@g.bracu.ac.bd).}%
\thanks{Saifur Rahman Sabuj is with the Department of Electrical and Electronic Engineering, BRAC University, Dhaka, Bangladesh, and with the Department of Automotive Engineering, Hanyang University, Seoul, Republic of Korea (e-mail: s.r.sabuj@ieee.org).}%
}

\maketitle

\begin{abstract}
Photoplethysmography (PPG)-based heart-rate (HR) estimation on smartphones remains unreliable in free-living conditions because motion artifacts vary in spectral structure across activities, and benchmarks such as BUT PPG v2.0 label signal quality only as binary good or bad, discarding partially usable data. This study develops an HR estimation architecture that explicitly conditions motion type and signal quality rather than treating both uniformly. We propose MoWaveQFormer, a three-stage architecture trained under ECG supervision. Stage 1 assigns each window a discrete motion group based on accelerometer-derived spectral energy, without trainable parameters. Stage 2 uses this index to select one of the three learnable FIR filter banks for motion-specific spectral shaping of the PPG signal. Stage 3 embeds the resulting sub-bands into patch tokens, re-weights them via a differentiable soft gate derived from the quality label, and encodes them with a Transformer whose pooled output is regressed to HR, trained jointly with an ECG-supervised loss and a pulse-transit-time consistency term. In a subject-independent split of BUT PPG v2.0 (3,888 recordings, 50 subjects), MoWaveQFormer achieved a mean absolute error of 7.85~bpm, the lowest among five methods, with significant improvements over three baselines (Wilcoxon test, $p<0.05$). Ablation and Bland-Altman analyzes characterize each component's contribution. With 816,445 parameters and sub-3-ms latency, MoWaveQFormer suits real-time deployment, pending validation on smartphone hardware. Replacing motion-agnostic filtering and binary quality discarding with differentiable conditioned processing offers a compact pathway to more reliable free-living PPG-based cardiovascular monitoring.
\end{abstract}
\begin{IEEEkeywords}
Deep learning, heart rate estimation, motion artifact removal, photoplethysmography, Transformer, wearable sensors, wavelet filter bank.
\end{IEEEkeywords}

\section{Introduction}
\label{sec:introduction}

\IEEEPARstart{C}{ontinuous} cardiovascular monitoring in free-living environments has emerged as a cornerstone of personalized medicine and preventive healthcare \cite{ref1,ref2}. Wearable optical sensors embedded in smartphones, smartwatches, and fitness trackers enable unobtrusive acquisition of photoplethysmographic (PPG) signals, making remote HR monitoring accessible to billions of users worldwide \cite{ref3,ref4}. Unlike electrocardiography, which requires dedicated electrodes, PPG leverages the principles of light absorption and reflection to detect blood volume oscillations in peripheral tissues, making it an attractive solution for mass-market deployment \cite{ref5}. However, the clinical accuracy and reliability of wearable PPG-based HR estimation remain challenged by the inherent fragility of the signal acquisition process in real-world, uncontrolled settings.

Motion artifacts stand for the dominant source of signal degradation in wearable PPG systems. Unlike laboratory or hospital environments where subjects are at rest or moving predictably, free-living acquisition involves heterogeneous physical activities that produce distinct spectral patterns of corruption \cite{ref6,ref8}. Accelerometer signals routinely captured alongside PPG remain underutilized by contemporary methods: many existing models either ignore acceleration entirely \cite{ref12,ref13} or treat it as a generic, motion-agnostic reference channel without explicitly exploiting its diagnostic value for distinguishing motion-induced artifacts \cite{ref14}.

Classical signal-processing approaches employ adaptive filtering, wavelet decomposition, and spectral-domain analysis, often paired with accelerometer-referenced correction schemes \cite{ref9,ref10}. While computationally efficient and interpretable, these methods rely on fixed decomposition bases and globally optimized parameters that degrade when artifact and physiological signal spectra overlap \cite{ref7,ref11}. Deep learning approaches have shifted the paradigm toward data-driven architectures: CNN and LSTM-based regressors achieve substantial error reductions on standard benchmarks \cite{ref12,ref13}, and recent Transformer-based approaches exploit self-attention to model motion dependencies without hand-crafted feature engineering \cite{ref14,ref15}. Yet despite these advances, two critical limitations persist across the literature.

First, existing deep models, whether CNN, LSTM, or Transformer-based, typically learn a single, globally optimal transformation that is applied uniformly to all input windows, irrespective of the underlying source of corruption \cite{ref13,ref14}. This architecture implicitly assumes that a learned filter bank suitable for stride artifacts is equally effective for coughing or talking, an assumption that contradicts the known, activity-specific spectral signatures of wearable PPG degradation \cite{ref8}. Knowledge-informed learning paradigms have begun to address this limitation by embedding domain priors into deep pipelines \cite{ref16}, but existing implementations do not explicitly condition their processing on a verified, externally-labeled motion category derived from accelerometer data. Second, signal-quality assessment in public PPG datasets remains fundamentally binary: the Brno University of Technology Smartphone PPG Database (BUT PPG) v2.0 dataset \cite{refBUTPPGv2}, widely used as a benchmark, labels each recording as either good or bad \cite{ref17}. This forced dichotomy compels downstream HR models into a hard accept/discard decision, wasting the partial informational content of borderline-quality segments and discarding physiological data that a soft, differentiable quality weighting could partially recover \cite{ref18}. Uncertainty-aware models have attempted to quantify prediction confidence
through belief propagation \cite{ref18} and Bayesian uncertainty estimation
\cite{ref19,ref20}; however, these approaches treat uncertainty primarily as a post-hoc output rather than an internal, differentiable signal that
modulates the encoder during both training and inference.

A third, often-overlooked gap concerns the training objective itself. While most contemporary PPG HR estimators are supervised directly on ECG-derived gold-standard HR labels, very few exploit additional physiological constraints such as PTT-the temporal delay between the R-peak of the ECG and the peak of the PPG waveform, which encodes information about arterial compliance and should remain consistent across short windows \cite{ref28}. Foundation-model approaches such as SiamQuality have demonstrated that self-supervised pretraining on massive unlabeled PPG corpora can confer quality-robustness, but this implicit learning of robustness does not leverage explicit, interpretable motion information or physiological supervision from synchronized ECG recordings \cite{ref22}.

Table~\ref{tab:limitations} compares five representative papers and states how each documented limitation motivates a specific MoWaveQFormer component.

\begin{table}[h]
\caption{Related Works and MoWaveQFormer Solutions}
\label{tab:limitations}
\centering
\scriptsize
\renewcommand{\arraystretch}{1}
\begin{tabular}{|p{1.6cm}|p{2.5cm}|p{2.5cm}|}
\hline
\textbf{Paper} & \textbf{Limitation} & \textbf{MoWaveQFormer Solution} \\
\hline
KID-PPG \cite{ref23} & Single filter for all motions; hard quality gating & Motion-aware filtering + soft quality attention \\
\hline
SiamQuality \cite{ref22} & Implicit robustness; no motion/ECG information & Explicit motion + ECG-PTT supervision \\
\hline
Cross-Attn PPG \cite{ref14} & Implicit motion learning; no quality gating & Explicit motion classification + quality gate \\
\hline
BeliefPPG \cite{ref18} & Uncertainty output & Quality integrated into attention mechanism \\
\hline
VMD/Wavelet \cite{ref10,ref11} & Motion-agnostic parameters & Learnable motion-conditioned filtering \\
\hline
\end{tabular}
\end{table}

To address these gaps, we introduce MoWaveQFormer, a motion-stratified, wavelet-guided transformer comprising three stages: a rule-based motion conditioner, a motion-conditioned wavelet filter bank, and a quality-gated Transformer encoder. Our main contributions are:
\begin{enumerate}
\item \textbf{Stage 1 (Motion Conditioning):} We propose MoWaveQFormer, a motion-conditioned framework for PPG heart-rate estimation that organizes motion into three spectral corruption profiles (subtle/rest, walking, and burst) and employs a learnable motion-specific wavelet filter bank for adaptive feature extraction. For the experiments reported in this study, motion groups are assigned from the recorded activity annotations to isolate the contribution of the downstream architecture, while the proposed accelerometer-based rule classifier is evaluated separately as a candidate for future real-time deployment.
\item \textbf{Stage 2--3 (Filtering and Quality Gating):} We develop a quality-gated transformer mechanism that incorporates signal reliability into feature learning without discarding low-quality segments, enabling the use of all 3,888 recordings, including 3,058 (78.7\%) annotated as poor quality.
\item \textbf{Training Objective:} We incorporate a differentiable PTT consistency constraint together with ECG-based HR supervision, and validate the proposed framework through extensive experiments and ablation studies on BUT PPG v2.0. MoWaveQFormer establishes the first systematic per-motion-class HR benchmark across all eight activity conditions and demonstrates statistically significant improvements over three of four representative motion-agnostic baselines (Wilcoxon signed-rank test, $p<0.05$), while showing no statistically significant difference from a ResNet1D baseline inspired by the Q-PPG architecture \cite{ref13}.
\end{enumerate}

\section{Methodology}
\label{sec:methodology}

\subsection{Overview of MoWaveQFormer Framework}
\label{subsec:overview}

\begin{figure}[t]
\centerline{\includegraphics[width=.5\textwidth]{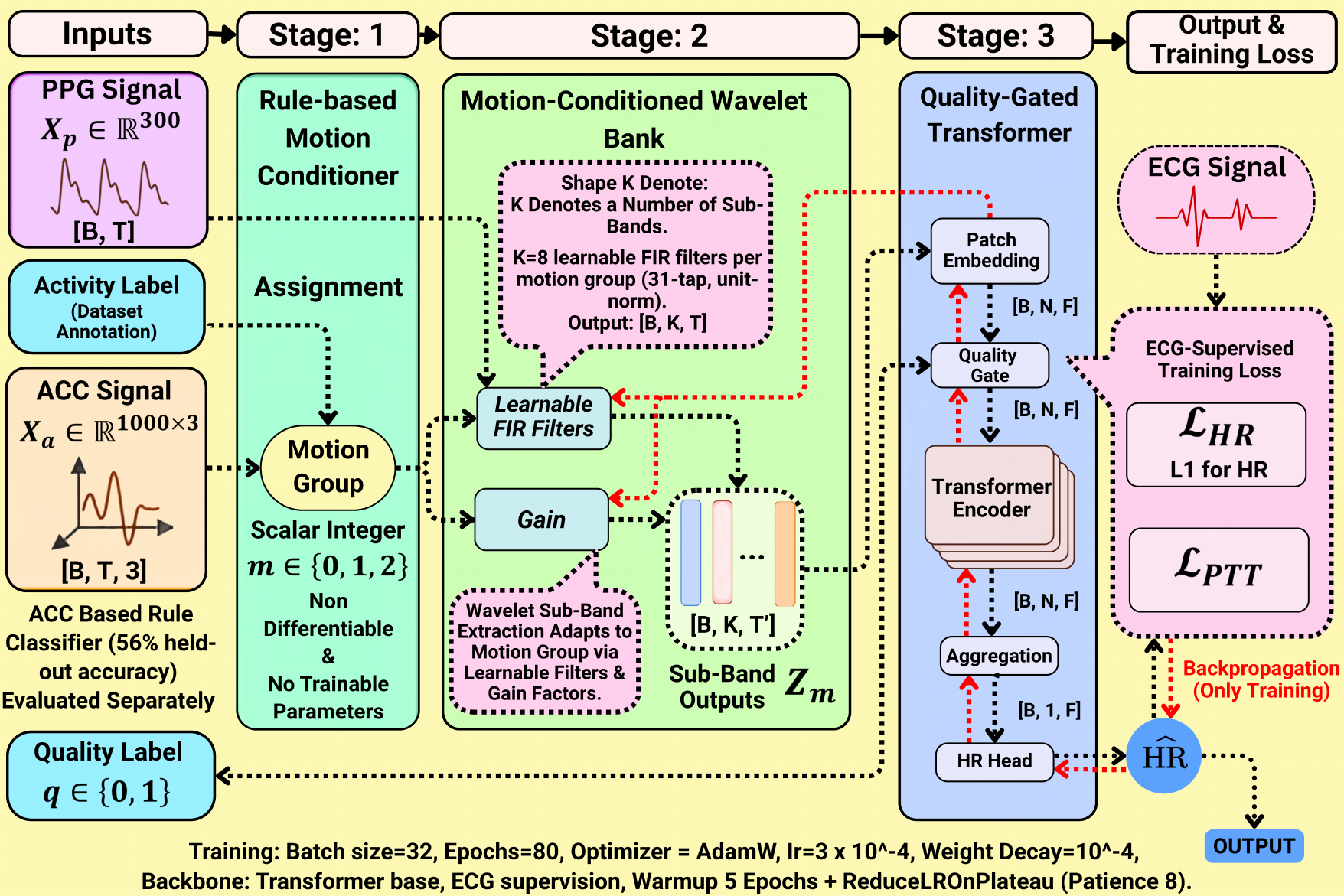}}
\caption{Overall architecture of MoWaveQFormer. A motion conditioner (Stage~1) supplies a scalar motion-group index; this index selects the motion-specific learnable filter bank in Stage~2, producing eight gated sub-band representations; Stage~3 embeds these sub-bands into patch tokens, applies a quality gate derived from the binary SQI label, and encodes the gated tokens with a Transformer encoder before aggregation and HR regression. Training combines an ECG-supervised heart-rate loss and a PPT consistency loss, $\mathcal{L}_{\mathrm{total}}=\mathcal{L}_{\mathrm{HR}}+\lambda_{\mathrm{PTT}}\mathcal{L}_{\mathrm{PTT}}$, with $\lambda_{\mathrm{PTT}}=0.1$.}
\label{fig2}
\end{figure}

Figure.~\ref{fig2} presents the overall architecture of MoWaveQFormer, a three-stage pipeline that maps a synchronized triplet of smartphone-acquired signals to a single heart-rate estimate. Formally, given a 10-second PPG segment $\mathbf{X}_p \in \mathbb{R}^{T}$ ($T=300$ at 30~Hz) and a concurrent tri-axial accelerometer segment $\mathbf{X}_a \in \mathbb{R}^{T_a \times 3}$ ($T_a=1000$ at 100~Hz), the network learns a function

\begin{equation}
\mathrm{\widehat{HR}} = f_\theta(\mathbf{X}_p, m, q),
\label{eq:mapping}
\end{equation}

\noindent where $m \in \{0,1,2\}$ is a discrete motion group and $q \in \{0,1\}$ is the binary signal-quality label associated with the window. In principle $m$ is obtained from the accelerometer signal via the rule-based conditioner described in Section~\ref{subsec:motion}; in the experiments reported in this paper, $m$ is instead taken directly from the recorded activity annotation for each window, with the accelerometer-derived classifier evaluated separately as a candidate for annotation-free deployment (Section~\ref{subsec:motion}). The network is trained under the supervision of ECG-derived heart-rate labels and R-peak annotations, the latter used only during training to compute the PTT target.

The pipeline proceeds in three stages, illustrated left-to-right in Figure~\ref{fig2}: (1) \textbf{Stage 1}: a motion conditioner assigns a discrete motion-group index $m$; (2) \textbf{Stage 2}: a motion-conditioned wavelet bank transforms $\mathbf{X}_p$ into $K=8$ motion-specific sub-band representations $\mathbf{Z}_m \in \mathbb{R}^{B \times 8 \times 300}$; and (3) \textbf{Stage 3}: a quality-gated Transformer encoder converts $\mathbf{Z}_m$ into a pooled latent representation while re-weighting patch tokens according to $q$, and a lightweight regression head produces $\mathrm{\widehat{HR}}$, jointly optimized against an ECG-supervised regression loss and a PTT-consistency term. Unlike architectures that concatenate accelerometer or motion features into a shared embedding, MoWaveQFormer uses $m$ as an explicit conditioning variable that selects which learned filters are applied to the PPG waveform before any feature encoding takes place, so that motion information shapes representation learning from the earliest processing step rather than being fused only near the output.

All PPG channels are independently min-max normalized to $[-1,1]$ within each window,
\begin{equation}
x' = \frac{2(x-x_{\min})}{x_{\max}-x_{\min}} - 1,
\label{eq:norm}
\end{equation}

\noindent where $x$ denotes either a PPG or an accelerometer channel. Applying
(\ref{eq:norm}) to the PPG and accelerometer signals yields the normalized
signals $X_p'$ and $X_a'$, respectively, which stabilizes the scale of the learnable filters in Stage~2 and the embedding projection in Stage~3 without requiring dataset-level statistics that could leak information across the subject-level split described in Section~\ref{sec:setup}.

\subsection{Motion Conditioning Module}
\label{subsec:motion}

Prior architectures typically treat accelerometer data as an auxiliary input stream rather than as a variable that actively conditions downstream processing. MoWaveQFormer instead computes an explicit motion descriptor

\begin{equation}
m = f_\phi(\mathbf{X}_a'), \qquad m \in \{0,1,2\},
\label{eq:motion}
\end{equation}

\noindent where $f_\phi$ is a deterministic, rule-based classifier rather than a learned network, so $\phi$ denotes fixed thresholds rather than trainable weights. This design choice follows from the activity distribution in BUT PPG v2.0: coughing and laughing together account for 7.4\% of training windows (Section~\ref{sec:setup}), which we judged insufficient to train a reliable end-to-end motion classifier without risking overfitting to subject-specific idiosyncrasies rather than motion physics.\\

The classifier computes the accelerometer magnitude envelope 

\begin{equation}
\|\mathbf{a}_t\| =
\sqrt{a_{x,t}^{2}+a_{y,t}^{2}+a_{z,t}^{2}},
\end{equation}

\noindent removes its mean, and evaluates its power spectrum via FFT. Two spectral-energy ratios are then compared against fixed thresholds: the fraction of total energy in the 1--2.5~Hz stride band, and the ratio of energy in the 4--15~Hz burst band relative to the 0.5--4~Hz baseline band. A window is assigned to Group~1 (walking) if the stride-band fraction exceeds 0.35 and the absolute stride-band energy exceeds a fixed floor, which prevents low-energy segments from triggering the walking branch spuriously; otherwise it is assigned to Group~2 (burst, covering coughing and laughing) if the burst-to-baseline ratio exceeds a factor of two and the absolute burst-band energy exceeds a corresponding floor; all remaining windows default to Group~0 (subtle/rest), which subsumes rest, elevated sensor pressure, finger movement, lighting change, and talking. On held-out test-set accelerometer recordings, this classifier agrees with the annotated activity group in 56\% of cases, reflecting the known difficulty of separating subtle motions (pressure, finger movement, talking) using accelerometer statistics alone. The resulting scalar $m$ carries no gradient and contributes no trainable parameters; it is used verbatim as an index into the filter bank of Stage~2.

\subsection{Motion-Conditioned Learnable Wavelet Filtering}
\label{subsec:wavelet}

Fixed-basis wavelet decomposition and globally optimized adaptive filters implicitly assume that a single set of decomposition parameters is appropriate across all motion conditions; accelerometer-referenced multistage cancellation schemes have shown that decomposing the acceleration reference itself into activity-dependent modes already improves artifact suppression over a single fixed reference model \cite{ref24}, and adaptive filters tuned specifically for exercise-induced corruption confirm that one globally optimized parameterization underperforms across heterogeneous motion regimes \cite{ref25}. This motivates replacing fixed decomposition with a bank of learnable finite-impulse-response (FIR) filters, indexed by motion group and jointly optimized with the HR regression objective rather than fitted through a separate signal-processing stage.

For each motion group $g \in \{0,1,2\}$, eight learnable filters $\mathbf{h}_{g,k} \in \mathbb{R}^{31}$, $k=1,\ldots,8$, are maintained as trainable parameters. Given the motion index $m$ for a record, the corresponding filter set $\mathbf{h}_{m,\cdot}$ is applied to the normalized PPG waveform via zero-padded convolution, and each resulting sub-band is scaled by a learned, motion-specific sigmoid gain:
\begin{equation}
\mathbf{Z}_m(t) =
\sum_{k=1}^{8}
\big[\mathbf{h}_{m,k} * \mathbf{X}_p'\big](t)
\cdot \sigma(w_{m,k}),
\label{eq:wavelet}
\end{equation}

\noindent where $*$ denotes convolution with padding of 15 samples on each side to preserve the input length, and $w_{m,k}$ are learned gain parameters distinct for every group-filter pair. The filter bank contains $3 \times 8 \times 31 = 744$ convolutional weights plus 24 gain parameters, all optimized end-to-end; only the group index $m$ is externally supplied rather than learned. The output $\mathbf{Z}_m \in \mathbb{R}^{8 \times T}$ replaces the raw PPG waveform as the input to the encoder in Stage~3.

\subsection{Quality-Guided Transformer Representation Learning}
\label{subsec:transformer}

The sub-band stack $\mathbf{Z}_m \in \mathbb{R}^{8 \times 300}$ is reshaped into 30 non-overlapping temporal patches of width 10 samples, flattened to vectors of dimension $8 \times 10 = 80$, and linearly projected to a model dimension of 128:
\begin{equation}
\mathbf{e}_i =
\mathbf{W}_{\mathrm{emb}}
\cdot
\mathrm{flatten}
\big(
\mathbf{Z}_m[:,\,10i:10(i{+}1)]
\big),
\quad i=1,\ldots,30,
\label{eq:embed}
\end{equation}

\noindent with $\mathbf{W}_{\mathrm{emb}} \in \mathbb{R}^{128\times80}$, after which a learnable positional encoding of matching dimension is added to every patch token.

Rather than discarding windows flagged as low quality by the binary SQI label, MoWaveQFormer converts the label into a continuous gate that re-weights, but never zeroes out, each patch token. The binary quality value is passed through a learned linear layer, broadcast across all 30 patches, squashed with a sigmoid, and applied multiplicatively to the patch embeddings:
\begin{equation}
\mathbf{g}_i =
\sigma
\big(
\mathbf{W}_q q + b_q
\big)_i
\odot
\mathbf{e}_i,
\qquad i=1,\ldots,30,
\label{eq:qgate}
\end{equation}

\noindent where $\mathbf{W}_q \in \mathbb{R}^{30\times1}$, $b_q\in\mathbb{R}^{30}$, and $\odot$ denotes element-wise scaling of each token by its gate value. This gating is applied to the embeddings entering the encoder rather than to the attention weights directly, so a low-quality token still participates in self-attention as a key and value at reduced magnitude rather than being masked out; the design reflects that a large majority of records in BUT PPG v2.0 carry the low-quality label (Section~\ref{sec:setup}), where a hard-discard policy would eliminate most of the available supervision.

The gated tokens $\{\mathbf{g}_i\}_{i=1}^{30}$ pass through a pre-norm Transformer encoder with 4 layers, 4 attention heads, model dimension 128, feed-forward dimension 512, and dropout 0.1, using standard scaled dot-product self-attention

\begin{equation}
\mathbf{Q}=\mathbf{G}\mathbf{W}_Q,\qquad
\mathbf{K}=\mathbf{G}\mathbf{W}_K,\qquad
\mathbf{V}=\mathbf{G}\mathbf{W}_V,
\label{eq:qkv}
\end{equation}

\noindent where
$\mathbf{G}=[\mathbf{g}_1,\ldots,\mathbf{g}_{30}]^{\top}
\in\mathbb{R}^{30\times128}$ denotes the matrix of gated patch tokens from
Eq.~\eqref{eq:qgate}, and
$\mathbf{W}_Q,\mathbf{W}_K,\mathbf{W}_V\in\mathbb{R}^{128\times128}$
are learnable projection matrices of the self-attention layer.

\begin{equation}
\mathbf{A}
=
\mathrm{softmax}
\left(
\frac{\mathbf{Q}\mathbf{K}^{\top}}
{\sqrt{d}}
\right);\label{eq:soft}
\end{equation}

\noindent because gating is applied upstream of the encoder, no additional masking is introduced inside the attention computation. The 30 output tokens are aggregated by global average pooling into a single 128-dimensional feature vector, passed through a two-layer regression head (128$\rightarrow$64 with GELU activation and dropout 0.1, then 64$\rightarrow$1) to produce $\mathrm{\widehat{HR}}$.

\subsection{Physiological Constraint and Training Objective}
\label{subsec:loss}

Beyond direct ECG-derived HR supervision, MoWaveQFormer incorporates a differentiable constraint based on PTT, the delay between the ECG R-peak and the corresponding PPG waveform peak, which reflects arterial compliance and should vary smoothly with vessel stiffness rather than erratically across a short window \cite{ref28}. For each window, a measured target $\mathrm{PTT}_{\mathrm{meas}}$ is computed as the mean delay between each detected ECG R-peak and the first subsequent PPG peak, restricted to a physiologically plausible interval of 50--600~ms; windows for which no valid match satisfies this interval are excluded only from the PTT term via a validity indicator, while HR supervision proceeds unaffected for those windows.

Given the network's HR output, an expected PTT is derived through a fixed physiological scaling relationship,

\begin{equation}
\widehat{\mathrm{PTT}}
=
\frac{60}{\mathrm{\widehat{HR}}}
\times
0.35,
\label{eq:ptt_pred}
\end{equation}

\noindent and the consistency loss penalizes the absolute discrepancy between the predicted and measured delay for windows with a valid target:

\begin{equation}
\mathcal{L}_{\mathrm{PTT}}
=
\left|
\widehat{\mathrm{PTT}}
-
\mathrm{PTT}_{\mathrm{meas}}
\right|.
\label{eq:ptt_loss}
\end{equation}

The overall training objective combines an L1 regression loss on the ECG-derived reference heart rate, $\mathcal{L}_{\mathrm{HR}}$, with the PTT consistency term:

\begin{equation}
\mathcal{L}_{\mathrm{total}}
=
\mathcal{L}_{\mathrm{HR}}
+
\lambda_{\mathrm{PTT}}
\mathcal{L}_{\mathrm{PTT}},
\label{eq:total}
\end{equation}

\noindent with $\lambda_{\mathrm{PTT}}=0.1$ held fixed throughout training. The PTT term introduces no new trainable parameters; it acts purely as a regularizer discouraging heart-rate predictions inconsistent with the arterial-delay characteristics observed in the paired ECG-PPG recording. The training procedure is summarized in Algorithm~\ref{alg:training}.

\begin{algorithm}[t]
\caption{Training procedure of MoWaveQFormer}
\label{alg:training}
\begin{algorithmic}[1]
\STATE \textbf{Input:} $\mathbf{X}_p$, $m$, $q$, $\mathrm{HR}_{\mathrm{ECG}}$, $\mathrm{PTT}_{\mathrm{meas}}$; learning rate $\eta$; loss weight $\lambda_{\mathrm{PTT}}$; max epochs $E_{\max}$
\STATE \textbf{Initialize:} Epoch $t \leftarrow 0$; parameters $\theta = \{\mathbf{h}_{g,k}, w_{g,k}, \mathbf{W}_{\mathrm{emb}}, \mathbf{W}_q, \mathbf{b}_q, \mathbf{W}^Q,\mathbf{W}^K,\mathbf{W}^V, \text{encoder, head}\}$ 
\REPEAT
    \FOR{each mini-batch}
        \STATE \textbf{Forward Propagation:}
        \STATE \hspace{1em} Normalize $\mathbf{X}_p$ with Eq. (\ref{eq:norm})
        \STATE \hspace{1em} $\mathbf{Z}_m \leftarrow$ MotionWaveletBank$(\mathbf{X}_p, m)$ with Eq. (\ref{eq:wavelet})
        \STATE \hspace{1em} $\{\mathbf{e}_i\}\leftarrow$ PatchEmbed$(\mathbf{Z}_m)$ with Eq. (\ref{eq:embed})
        \STATE \hspace{1em} $\{\mathbf{g}_i\}\leftarrow$ QualityGate$(\{\mathbf{e}_i\},q)$ with Eq. (\ref{eq:qgate})
        \STATE \hspace{1em} $\mathbf{f}\leftarrow$ GlobalPool(Transformer$(\{\mathbf{g}_i\})$) with Eqs. (\ref{eq:qkv})--(\ref{eq:soft})
        \STATE \hspace{1em} $\mathrm{\widehat{HR}}\leftarrow$ RegressionHead$(\mathbf{f})$
        \STATE \textbf{Loss Computation:}
        \STATE \hspace{1em} Compute total loss:
            $\mathcal{L}_{\mathrm{total}} = |\mathrm{\widehat{HR}}-\mathrm{HR}_{\mathrm{ECG}}| + \lambda_{\mathrm{PTT}}\,\mathcal{L}_{\mathrm{PTT}}$
        \STATE \textbf{Backpropagation:}
        \STATE \hspace{1em} Evaluate gradient $\nabla_\theta \mathcal{L}_{\mathrm{total}}$ \COMMENT{no gradient for $m$}
        \STATE \hspace{1em} Update $\theta$ with AdamW
    \ENDFOR
    \STATE Validate; update $\eta$; $t\leftarrow t+1$
\UNTIL{$t \geq E_{\max}$ or validation MAE converges}
\STATE \textbf{Output:} Trained parameters $\theta$
\end{algorithmic}
\end{algorithm}

\section{Experimental Setup}
\label{sec:setup}

\subsection{Dataset and Preprocessing}
\label{subsec:dataset}

MoWaveQFormer is evaluated on the BUT PPG v2.0 database, comprising 3,888 ten-second recordings from 50 subjects (25 male, 25 female; age 33.9 $\pm$ 16.7 years) across eight activity conditions: rest, elevated sensor pressure, finger movement, walking, coughing, laughing, lighting change, and talking. Each record provides smartphone PPG at 30~Hz (300 samples), a reference ECG at 1000~Hz, and, for 3,840 of the 3,888 records, a synchronized tri-axial chest accelerometer at 100~Hz. Each window carries a binary signal-quality index (SQI); across the full dataset, 3,058 windows (78.7\%) are labeled low quality and 830 (21.3\%) high quality, reflecting the class imbalance typical of free-living acquisition and motivating the soft quality gate of Section~\ref{subsec:transformer}. The reference heart rate for each window is the single annotated value provided per record; PPG channels are normalized as in Eq.~(\ref{eq:norm}). During training, each sample independently receives additive Gaussian noise ($\sigma=0.02$) with probability 0.5 and, independently, a random circular temporal shift of up to $\pm10$ samples with probability 0.5; no augmentation is applied at validation or test time.

\subsection{Data Partitioning}
\label{subsec:split}

The dataset is partitioned at the subject level to prevent identity leakage across splits: 35 subjects (2,742 recordings, 70.5\%) form the training set, 7 subjects (548 recordings, 14.1\%) form the validation set, and the remaining 8 subjects (598 recordings, 15.4\%) form the test set, using a fixed random seed. Within the training set, the eight activity labels collapse into markedly imbalanced motion groups under the mapping of Section~\ref{subsec:motion}: 2,436 windows (88.9\%) fall into the subtle/rest group, 102 (3.7\%) into walking, and 204 (7.4\%) into the burst group. This imbalance, present identically in the test set (532, 22, and 44 windows for the three groups, respectively), is referenced throughout the evaluation to contextualize per-group results.

\subsection{Evaluation Metrics}
\label{subsec:metrics}

Performance is measured against the ECG-derived reference heart rate using mean absolute error (MAE), root-mean-square error (RMSE), and Pearson correlation coefficient $r$ on the test set. Results are additionally stratified by SQI (good vs. poor quality) to assess the value of the quality gate, and by each of the eight annotated activity classes to characterize motion-specific performance. Statistical comparisons between MoWaveQFormer and each baseline are conducted using a two-sided Wilcoxon signed-rank test on paired per-record absolute errors, with all methods evaluated on an identical, record-matched subset of the test set; effect size is reported as $r = |Z|/\sqrt{n}$. Agreement with the ECG reference is further characterized using Bland-Altman analysis, reporting mean bias and 95\% limits of agreement ($\text{bias} \pm 1.96\,\text{SD}$).

\subsection{Training Configuration}
\label{subsec:training}

MoWaveQFormer and all ablated variants are trained end-to-end using AdamW with an initial learning rate of $3\times10^{-4}$ and weight decay of $10^{-4}$, for 80 epochs with a batch size of 32. The learning rate follows a linear warmup over the first 5 epochs, followed by a ReduceLROnPlateau schedule (factor 0.5, patience 8 epochs) monitored on validation MAE. Random seeds are fixed for Python, NumPy, and PyTorch, with deterministic algorithms enabled where supported by the underlying kernels. The complete model comprises the motion conditioner (zero trainable parameters), the motion-conditioned wavelet bank (768 parameters), and the quality-gated Transformer encoder together with embedding, positional-encoding, and regression-head layers (approximately 815,700 parameters), totaling 816,445 trainable parameters. All training and evaluation is performed on a cloud-hosted NVIDIA GPU (Kaggle compute environment). Inference latency for a single 10-second window was measured over 200 runs following a 50-iteration warmup, on both GPU and CPU (Section~\ref{subsec:efficiency}).

\section{Results}
\label{sec:results}

\subsection{Heart-Rate Estimation Accuracy}
\label{subsec:overall}

MoWaveQFormer is compared against five representative baselines spanning classical spectral tracking and deep regression. \textbf{TROIKA} \cite{ref8} bandpass-filters the PPG signal to the 0.67--3~Hz heart-rate band and reports the dominant FFT peak as the HR estimate. A stronger spectral-tracking baseline, denoted \textbf{Harmonic-SP}, extends this approach with harmonic reinforcement and continuity-constrained frequency tracking following Sch\"ack et al.~\cite{refTAPIR}. \textbf{DeepPPG} follows the convolutional regression architecture of Reiss et al.~\cite{ref12}. \textbf{CNN-BiLSTM} combines convolutional feature extraction with bidirectional LSTM layers. \textbf{ResNet1D} follows a Q-PPG-style residual architecture \cite{ref13}, with three residual blocks of increasing channel width interleaved with strided convolutions. All learned baselines are trained on the same training partition using identical augmentation, optimization, and evaluation protocols.

\begin{table}[h]
\caption{Overall HR Estimation on the Test Set (ECG Reference, $n=598$)}
\label{tab:overall}
\centering
\scriptsize
\begin{tabular}{|l|c|c|c|}
\hline
\textbf{Method} & \textbf{MAE (bpm)} & \textbf{RMSE (bpm)} & \textbf{Pearson $r$} \\
\hline
TROIKA \cite{ref8}            & 20.559 & 26.129 & 0.124 \\
\hline
Harmonic-SP \cite{refTAPIR}   & 21.497 & 26.686 & 0.180 \\
\hline
DeepPPG \cite{ref12}          & 8.430  & 13.846 & 0.152 \\
\hline
CNN-BiLSTM                    & 8.479  & 14.015 & 0.067 \\
\hline
ResNet1D \cite{ref13}         & 8.175  & 14.246 & 0.101 \\
\hline
\textbf{MoWaveQFormer (Ours)}     & \textbf{7.851} & \textbf{13.377} & \textbf{0.292} \\
\hline
\end{tabular}
\end{table}

Table~\ref{tab:overall} reports overall test-set performance ($n=598$) for MoWaveQFormer and the five baselines. Both classical spectral-tracking methods, TROIKA \cite{ref8} and Harmonic-SP \cite{refTAPIR}, exhibit MAE exceeding 20~bpm, confirming that fundamental-frequency tracking alone is inadequate under the heterogeneous motion conditions present in BUT PPG v2.0. All three deep learning baselines substantially outperform the spectral trackers, with MAE in the 8.2--8.5~bpm range. MoWaveQFormer achieves the best overall performance among all six methods, with the lowest estimation error and highest correlation with the ECG reference (Table~\ref{tab:overall}), outperforming the strongest baseline, ResNet1D \cite{ref13}.

\begin{table}[h]
\caption{Wilcoxon Signed-Rank Test: MoWaveQFormer vs. Baselines ($n=598$)}
\label{tab:wilcoxon}
\centering
\scriptsize
\begin{tabular}{|l|c|c|c|}
\hline
\textbf{Method} & \textbf{$p$-value} & \textbf{Effect size $r$} & \textbf{Significant} \\
\hline
vs. Harmonic-SP  & $3.1\times10^{-69}$ & 0.719 & Yes \\
\hline
vs. DeepPPG      & 0.017 & 0.098 & Yes \\
\hline
vs. CNN-BiLSTM   & 0.016 & 0.099 & Yes \\
\hline
vs. ResNet1D     & 0.319 & 0.041 & No \\
\hline
\end{tabular}
\end{table}

Statistical significance was assessed via a two-sided Wilcoxon signed-rank test on paired per-record absolute errors (Table~\ref{tab:wilcoxon}). TROIKA was excluded from formal significance testing given its MAE (20.6~bpm), already exceeding twice that of any learned baseline (Table~\ref{tab:overall}), making a paired comparison uninformative relative to the more competitive Harmonic-SP tracker. MoWaveQFormer's improvement over DeepPPG ($p = 0.017$), CNN-BiLSTM ($p = 0.016$), and Harmonic-SP ($p = 3.1 \times 10^{-69}$) is statistically significant, with a large effect size against the classical spectral tracker ($r = 0.719$) and small effect sizes against the two learned CNN baselines ($r \approx 0.10$). The improvement over ResNet1D did not reach statistical significance ($p = 0.319$, $r = 0.041$), despite MoWaveQFormer achieving the lower mean error; this comparison is examined further in Section~\ref{sec:discussion}.

\subsection{Motion-Specific Corruption Patterns and PTT Consistency}
\label{subsec:corruption}

\begin{figure}[h]
\centerline{\includegraphics[width=\columnwidth]{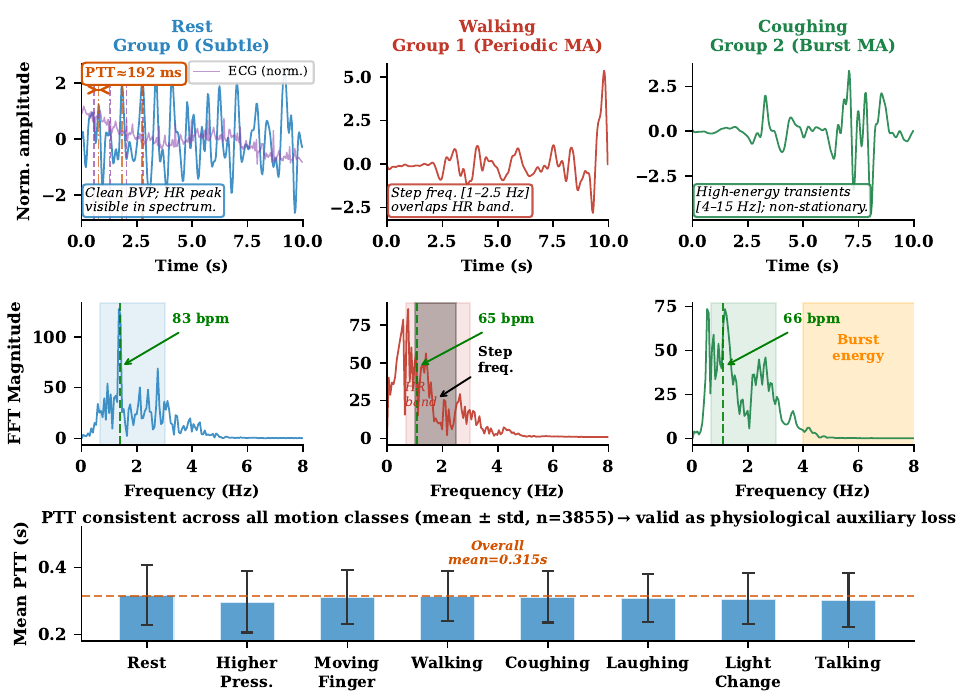}}
\caption{Motion-specific PPG corruption patterns and PTT measurement. Top: representative rest, walking, and coughing windows in the time domain, with ECG overlay and PTT annotation for the rest example. Middle: corresponding frequency spectra, with the annotated heart-rate peak, stride-frequency band, and burst-energy region indicated. Bottom: mean PTT ($\pm$SD) across all eight activity classes, showing consistency across motion conditions.}
\label{fig3}
\end{figure}

Figure~\ref{fig3} illustrates the qualitative basis for the motion-group taxonomy used in Section~\ref{subsec:motion}. Representative rest, walking, and coughing windows show visually distinct time- and frequency-domain corruption signatures: the rest window exhibits a clean, periodic waveform with a single dominant spectral peak coinciding with the ECG-derived reference rate, the walking window shows broadband energy concentrated in the 1--2.5~Hz stride band overlapping the heart-rate band, and the coughing window shows a non-stationary, transient burst that does not localize to a narrow frequency range. The bottom panel reports mean PTT, computed as the delay between ECG R-peaks and the corresponding PPG peak, across all eight activity classes ($n=3{,}855$ valid measurements). Mean PTT remains within a narrow range (0.303--0.318~s) across all classes, with overlapping standard deviations, supporting the use of a single fixed physiological scaling relationship (Eq.~(\ref{eq:ptt_pred})) as a training-time constraint independent of activity type.

\subsection{Effect of the Quality Gate}
\label{subsec:qualityresults}

\begin{table}[h]
\caption{MAE (bpm) Stratified by Signal Quality}
\label{tab:quality}
\centering
\scriptsize
\begin{tabular}{|l|c|c|}
\hline
\textbf{Method} & \textbf{Low quality (SQI=0)} & \textbf{High quality (SQI=1)} \\
\hline
TROIKA          & 24.690 & 6.522 \\
\hline
Harmonic-SP     & 25.759 & 7.018 \\
\hline
DeepPPG         & 9.340  & 5.339 \\
\hline
CNN-BiLSTM      & 9.274  & 5.776 \\
\hline
ResNet1D        & 9.386  & \textbf{4.060} \\
\hline
\textbf{MoWaveQFormer} & \textbf{8.805} & 4.609 \\
\hline
\end{tabular}
\end{table}

Table~\ref{tab:quality} reports MAE stratified by the binary signal-quality index (SQI). Across all methods, error is substantially higher on low-quality (SQI$=0$) windows, which constitute 3,058 of the 3,888 total recordings (78.7\%) \cite{ref17}. MoWaveQFormer achieves the lowest MAE on low-quality windows, whereas ResNet1D performs marginally better on high-quality windows (Table~\ref{tab:quality}). MoWaveQFormer's advantage is therefore concentrated in the low-quality regime that the soft quality gate Eq.~(\ref{eq:qgate}) was designed to address, rather than uniformly across all signal conditions. Qualitative inspection of representative good- and poor-quality windows confirms that heart-rate-relevant peak structure persists even in windows flagged as low quality by the binary SQI label, motivating the soft gating mechanism of Eq.~(\ref{eq:qgate}) over hard discarding.

\subsection{Per-Motion-Class Performance}
\label{subsec:permotion}

\begin{table*}[t]
\caption{Per-Motion-Class MAE (bpm) on the Test Set}
\label{tab:permotion}
\centering
\scriptsize
\begin{tabular}{|l|c|c|c|c|c|c|}
\hline
\textbf{Motion Class} & \textbf{TROIKA} & \textbf{Harmonic-SP} & \textbf{DeepPPG} & \textbf{CNN-BiLSTM} & \textbf{ResNet1D} & \textbf{MoWaveQFormer} \\
\hline
Rest              & 19.000 & 19.738 & 7.663  & 7.803  & 7.555  & \textbf{7.493} \\
\hline
Higher pressure   & 20.750 & 22.432 & \textbf{6.914}  & 6.985  & 7.807  & 7.499 \\
\hline
Moving finger     & 21.239 & 23.348 & \textbf{7.906}  & 8.226  & 9.769  & 9.918 \\
\hline
Walking           & 32.159 & 32.364 & 17.606 & 17.467 & 15.626 & \textbf{10.962} \\
\hline
Coughing          & 25.841 & 26.591 & 13.835 & 13.761 & 12.966 & \textbf{9.880} \\
\hline
Laughing          & 24.364 & 27.909 & 12.762 & 12.548 & 10.571 & \textbf{9.442} \\
\hline
Light change      & 23.830 & 24.693 & 7.381  & \textbf{6.779}  & 6.814  & 7.082 \\
\hline
Talking           & 22.250 & 23.523 & 8.347  & 8.229  & \textbf{6.820}  & 7.698 \\
\hline
\end{tabular}
\end{table*}

Table~\ref{tab:permotion} decomposes test-set MAE by the eight annotated activity classes. MoWaveQFormer achieves the lowest error among all methods for walking, coughing, and laughing (Table~\ref{tab:permotion}), corresponding to the walking and burst motion groups targeted explicitly by the motion-conditioned filter bank (Section~\ref{subsec:wavelet}). For rest, MoWaveQFormer is comparable to the best baseline (7.493 versus ResNet1D's 7.555). By contrast, for higher-pressure, moving-finger, light-change, and talking---the four remaining activities that, together with rest, are collapsed into the single subtle/rest motion group under the coarse three-way conditioning scheme of Section~\ref{subsec:motion}---MoWaveQFormer does not achieve the lowest error, being outperformed by DeepPPG on higher-pressure and moving-finger, by CNN-BiLSTM on light-change, and by ResNet1D on talking (Table~\ref{tab:permotion}).

\subsection{Ablation Study}
\label{subsec:ablationresults}

To isolate the contribution of each proposed component, three ablated variants of MoWaveQFormer are trained under an identical schedule to the full model (Section~\ref{subsec:training}). ($i$) \textbf{No motion wavelet}: Eq.~(\ref{eq:wavelet})'s per-group filter bank is replaced with a single shared filter set of equal parameter count. ($ii$) \textbf{No quality gate}: the gate in Eq.~(\ref{eq:qgate}) is fixed to unity for all tokens. ($iii$) \textbf{No PTT loss}: $\lambda_{\text{PTT}}=0$ in Eq.~(\ref{eq:total}). Each variant is trained independently and evaluated on the same validation and test partitions as the full model.

\begin{table}[h]
\caption{Ablation Study (Validation MAE, bpm)}
\label{tab:ablation}
\centering
\scriptsize
\begin{tabular}{|l|c|c|}
\hline
\textbf{Variant} & \textbf{Val. MAE (bpm)} & \textbf{$\Delta$ vs. full model} \\
\hline
Full MoWaveQFormer & 7.162 & --- \\
\hline
\makecell[l]{Shared filter bank\\(no motion conditioning)} & 7.079 & $-0.083$ \\
\hline
Uniform gate (no quality conditioning) & 7.283 & $+0.121$ \\
\hline
$\lambda_{\text{PTT}}=0$ (no PTT constraint) & 7.305 & $+0.143$ \\
\hline
\end{tabular}
\end{table}

Table~\ref{tab:ablation} reports validation-set MAE for the full model and each of the three ablated variants described above. Removing the quality gate increases validation MAE by 0.121~bpm, and removing the PTT-consistency term increases it by 0.143~bpm, indicating that both components contribute measurably to overall accuracy. Replacing the per-group filter bank with a single filter set shared across all motion groups changes validation MAE by only $-0.083$~bpm, a difference within the range of run-to-run variation observed across repeated seeded training runs in this work, and does not, on its own, indicate a discernible aggregate benefit from motion-specific filtering. This result is examined jointly with the per-motion-class findings of Section~\ref{subsec:permotion} and the filter-response analysis of Section~\ref{subsec:filteranalysis} in the Discussion.

\subsection{Learned Filter-Bank Analysis}
\label{subsec:filteranalysis}
\begin{figure}[h]
\centerline{\includegraphics[width=\columnwidth]{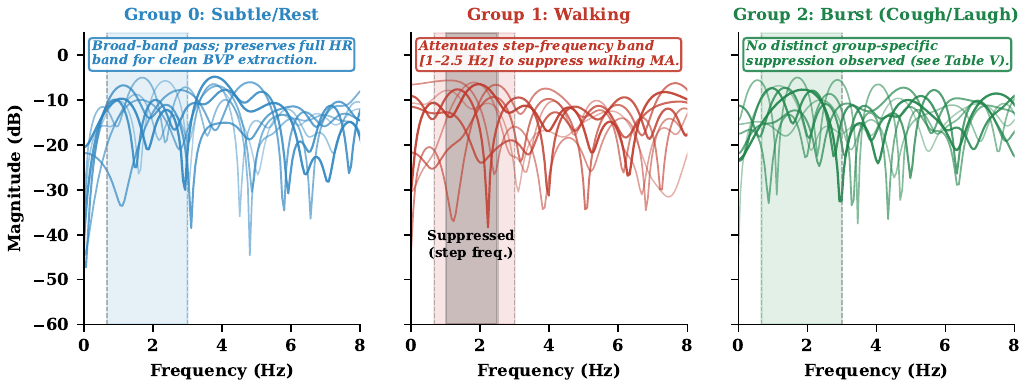}}
\caption{Frequency response of the eight learned FIR filters for each motion group, expressed in dB magnitude. The walking group (Group~1) shows modestly deeper attenuation within the stride-frequency band (1--2.5~Hz) relative to the subtle/rest group (Group~0); the burst group (Group~2) does not show comparable specialization within the burst-artifact band (4--8~Hz), consistent with the quantitative in-band magnitude comparison reported in the text and the ablation result in Table~\ref{tab:ablation}.}
\label{fig5}
\end{figure}
Figure~(\ref{fig5}) shows the frequency response of all eight learned filters for each of the three motion groups. Mean in-band magnitude, averaged across the eight filters within each group, is reported for three frequency bands of interest: the heart-rate band (0.67--3.0~Hz), the stride-frequency band (1.0--2.5~Hz), and the burst-artifact band (4.0--8.0~Hz, the portion of the nominal 4--15~Hz burst range within the plotted range). The walking group (Group~1) shows modestly greater attenuation in the stride-frequency band than the subtle/rest group (Group~0) ($-13.25$~dB versus $-12.96$~dB), consistent with a small degree of motion-specific specialization. The burst group (Group~2), by contrast, shows less attenuation in the burst-artifact band than Group~0 ($-14.09$~dB versus $-14.99$~dB), indicating that the filter bank did not learn a distinguishing suppression response for this group, despite its measurable per-class accuracy advantage reported in Table~\ref{tab:permotion}.This asymmetry likely reflects the differing temporal structure of the two artifact types: stride-induced motion during walking is approximately periodic within a 10-second window, producing a stationary spectral signature that a static FIR filter can represent as a persistent notch, whereas coughing and laughing introduce brief, non-stationary transients that a fixed-response filter cannot selectively suppress without also attenuating genuine cardiac information elsewhere in the window.

\subsection{Clinical Agreement Analysis}
\label{subsec:blandaltman}
\begin{figure}[h]
\centerline{\includegraphics[width=\columnwidth]{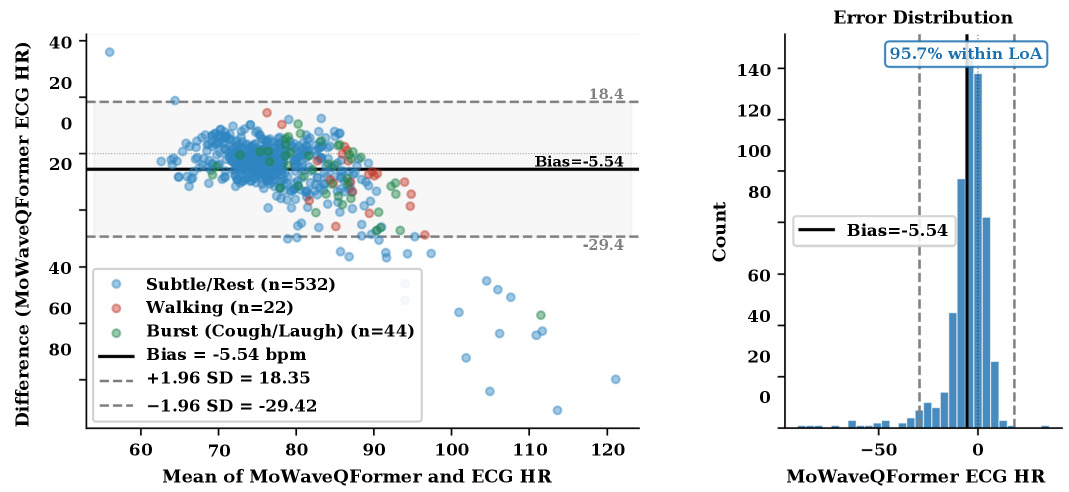}}
\caption{Bland-Altman agreement between MoWaveQFormer predictions and ECG reference HR on the test set ($n=598$), colour-coded by motion group. Bias $=-5.54$~bpm; 95\% limits of agreement $=[-29.42, 18.35]$~bpm; 95.7\% of errors fall within the limits of agreement. Right: histogram of prediction errors.}
\label{fig6}
\end{figure}
Figure~(\ref{fig6}) presents a Bland-Altman comparison of MoWaveQFormer predictions against the ECG reference on the test set ($n=598$). MoWaveQFormer shows a mean bias of $-5.54$~bpm (predictions systematically lower than the reference) with 95\% limits of agreement of $[-29.42, 18.35]$~bpm; 95.7\% of test-set errors fall within these limits. Inspection of the eleven largest errors (difference $<-50$~bpm) shows that all eleven occurred on windows flagged as low quality (SQI$=0$) with an annotated reference HR between 120 and 161~bpm, substantially above the dataset's overall mean HR of 78.2~$\pm$~13.3~bpm. For each of these eleven records, we additionally recomputed a reference HR directly from the dataset's provided QRS annotations via inter-beat interval averaging; in every case the recomputed value (65.8--84.5~bpm) was closer to MoWaveQFormer's prediction than the annotated CSV value was, with the CSV-to-recomputed ratio consistently near 2:1.

\subsection{Computational Efficiency}
\label{subsec:efficiency}

MoWaveQFormer contains 816,445 trainable parameters. Averaged over 200 forward passes following a 50-iteration warmup, inference on a single 10-second window required $2.04 \pm 0.11$~ms on a cloud-hosted CUDA GPU and $2.63 \pm 0.21$~ms on CPU (both measured on Kaggle compute infrastructure, batch size 1).

\section{Discussion}
\label{sec:discussion}

MoWaveQFormer's improvements over DeepPPG, CNN-BiLSTM, and the classical Harmonic-SP spectral tracker were statistically significant (Table~\ref{tab:wilcoxon}), indicating that explicit motion-conditioned feature extraction and quality-aware attention improve robustness under heterogeneous motion conditions \cite{ref8,ref28}. The improvement over ResNet1D, a residual architecture inspired by Q-PPG \cite{ref13}, did not reach significance. This pattern is best understood jointly with the per-class and ablation results rather than as an isolated outcome: MoWaveQFormer's clearest advantages occur for walking and the two burst-associated activities, coughing and laughing (Table~\ref{tab:permotion})---precisely the activities the motion conditioner separates into distinct groups---whereas for the four remaining subtle activities collapsed with rest into a single shared filter set, MoWaveQFormer shows no consistent advantage and is occasionally outperformed by baselines with no motion conditioning at all. This indicates the three-group taxonomy differentiates walking- and burst-type corruption from other activity but does not resolve heterogeneity within the subtle/rest group itself, where elevated pressure, finger movement, lighting change, and talking evidently produce distinct-enough corruption signatures that a single shared filter cannot address uniformly, consistent with prior characterizations of activity-specific PPG corruption spectra \cite{ref6,ref7,ref8}. The corresponding pooled ablation effect for motion conditioning (Table~\ref{tab:ablation}, $\Delta=-0.083$~bpm) is reconciled by the extreme class imbalance of the validation set, in which 88.9\% of windows belong to the subtle/rest group: a mechanism that meaningfully improves only the remaining 11.1\% will necessarily produce a small pooled-average effect while still yielding the substantial per-class gains reported for walking, coughing, and laughing specifically. The filter-response analysis (Figure~\ref{fig5}) offers an architecture-level explanation for why walking is specialized more clearly than burst motion: stride artifacts are approximately periodic within a 10-second window, producing a stationary spectral signature that a static FIR filter can represent as a persistent notch, whereas coughing and laughing introduce brief, non-stationary transients that a fixed-response filter cannot selectively suppress without also attenuating genuine cardiac information elsewhere in the same window---though the walking and burst groups remain minorities relative to the dominant subtle/rest class (88.9\%), which may still constrain gradient signal available to specialize either branch.

The quality-stratified results (Table~\ref{tab:quality}) and the corresponding ablation indicate that the soft quality gate provides a genuine, though modest, benefit concentrated on low-quality windows, where MoWaveQFormer achieves the lowest MAE of any evaluated method. Given that 78.7\% of BUT PPG v2.0 windows carry the low-quality label \cite{ref18}, a hard-discard policy would eliminate most of the available training signal; the differentiable re-weighting used here retains all windows while down-weighting unreliable segments, departing from uncertainty-aware approaches that treat reliability as a post-hoc network output \cite{ref19,ref20}. On high-quality windows, ResNet1D achieves marginally lower MAE, suggesting MoWaveQFormer's overall advantage is driven primarily by its behavior on the low-quality majority rather than uniform superiority. The PTT-consistency ablation showed the largest single-component effect of the three mechanisms (Table~\ref{tab:ablation}), supporting the fixed physiological HR--PTT scaling relationship~\cite{ref28} as a training-time regularizer.

All results reported here use the recorded activity annotation, rather than the accelerometer-derived rule-based classifier, to determine the motion group. On held-out test-set accelerometer data, the rule-based classifier agrees with the annotated activity group in only 56\% of cases, reflecting the known difficulty of discriminating subtle motions from rest using accelerometer statistics alone. The reported accuracy figures therefore represent an upper bound on what motion-conditioned filtering can achieve given accurate group assignment; deployment using the rule-based classifier in its current form would likely realize a smaller fraction of the reported walking- and burst-motion gains, since misclassification routes windows to a mismatched filter set. Bland-Altman analysis (Figure~\ref{fig6}) showed a mean bias of $-5.54$~bpm that increased in magnitude as true HR increased, most visibly above approximately 100~bpm, consistent with underrepresentation of elevated-HR samples in a training distribution with mean 78.2~$\pm$~13.3~bpm. The eleven largest errors were traced to low-quality windows whose annotated reference HR (120--161~bpm) disagreed by a factor of approximately two with an independently recomputed HR derived from the dataset's own QRS annotations, with MoWaveQFormer's predictions tracking the recomputed value more closely in every case, consistent with reference-label noise at extreme heart rates rather than systematic model error. At 816,445 parameters with 2.0~ms GPU and 2.6~ms CPU inference latency per window, MoWaveQFormer is substantially smaller than typical Transformer-based physiological signal models \cite{ref15} and compatible in principle with real-time operation, though these figures were obtained on cloud infrastructure rather than smartphone-class hardware.

These findings should be interpreted alongside several limitations. All results derive from a single subject-level split of one dataset; generalization to other populations, devices, and acquisition protocols remains unevaluated. The three-group motion taxonomy resolves walking- and burst-type corruption but not heterogeneity within the subtle/rest group, and the accelerometer classifier's 56\% held-out agreement limits applicability without ground-truth activity annotation; the heavy class imbalance toward subtle/rest (88.9\%) further constrains statistical power for evaluating motion-specific mechanisms. Repeated runs showed 0.1–0.2 bpm MAE variation from non-deterministic GPU kernels, causing the ResNet1D comparison's significance to vary across runs and warranting cautious interpretation. Finally, inference latency was benchmarked on cloud rather than on-device hardware, and reference labels for the extreme-outlier low-quality windows discussed above may themselves be unreliable, an issue inherent to the dataset rather than the proposed method.

Future work will pursue a finer-grained, clustering-derived motion taxonomy and continuously attention-gated filtering to better address subtle/rest heterogeneity and transient burst artifacts; a learned, class-balanced motion classifier to realize per-class gains without activity annotation; and cross-dataset, on-device validation to establish deployment feasibility beyond this single-dataset, cloud-benchmarked evaluation.

\section{Conclusion}
\label{sec:conclusion}
This paper introduced MoWaveQFormer, a three-stage framework for smartphone PPG heart-rate estimation that integrates motion-conditioned learnable wavelet filtering, a soft quality-gated Transformer encoder, and a PTT consistency constraint under ECG supervision. Evaluated on BUT PPG v2.0, MoWaveQFormer achieved the best overall performance among six representative methods, with statistically significant improvements over DeepPPG, CNN-BiLSTM, and a classical harmonic spectral tracker, while showing no significant difference from a Q-PPG-inspired ResNet1D baseline \cite{ref13}. The largest gains occurred during walking and burst-motion activities, consistent with the motion-conditioned filter bank's design objective, whereas more heterogeneous subtle-motion activities remained challenging under the current three-group taxonomy. Ablation and filter-response analyses showed that the soft quality gate and PTT-consistency constraint contributed consistently to performance, while motion-conditioned filtering benefited primarily the motion types it was designed to address. Bland--Altman analysis indicated good agreement with the ECG reference, with the largest discrepancies concentrated in low-quality, high-heart-rate recordings consistent with reference-label uncertainty rather than systematic model failure. Given its compact architecture and low inference latency, MoWaveQFormer is compatible in principle with real-time wearable deployment, and future work will target finer-grained motion characterization, learned real-time motion classification, cross-dataset validation, and on-device implementation.

\end{document}